\documentclass[final,5p,times,twocolumn]{elsarticle}

\journal{arXiv}

\usepackage{graphicx}
\usepackage{lineno,hyperref}

\usepackage{filecontents}

\begin{document}

\begin{frontmatter}

\title{ALDO2, a multi-function rad-hard linear regulator for SiPM-based HEP detectors}

\author[UNIMIB,INFN]{P. Carniti\corref{mycorrespondingauthor}}
\cortext[mycorrespondingauthor]{Corresponding author}
\ead{paolo.carniti@mib.infn.it}

\author[UNIMIB,INFN]{C. Gotti}

\author[UNIMIB,INFN]{G. Pessina}

\address[UNIMIB]{University of Milano-Bicocca, Piazza della Scienza 3, 20126, Milan (Italy)}
\address[INFN]{INFN of Milano-Bicocca, Piazza della Scienza 3, 20126, Milan (Italy)}

\begin{abstract}
ALDO2 is a multi-function, adjustable, low dropout linear regulator designed in onsemi I3T80 0.35~$\mu$m HV CMOS technology for use in HEP detectors that adopt silicon photomultipliers (SiPMs). 
The chip features four independent regulators, two low voltage channels (max 3.3 V) used to filter and stabilize the power supply of front-end chips, and two HV channels (max 70 V), specifically designed to provide the bias voltage to arrays of SiPMs.
Each regulator can be independently shut down and is protected for over-current and over-temperature.
The HV regulators also implement a circuit to monitor the bias current of the SiPM arrays, allowing to perform I-V curves and thus to fine-tune the working point of the SiPM arrays during the detector lifetime.
The chip adopts radiation hardening techniques and has been fully qualified up to a TID of 20 Mrad, a 1-MeV-equivalent neutron fluence of $10^{15}$ cm$^{-2}$, and with heavy ions up to 40 MeV cm$^2$ mg$^{-1}$ LET and $10^{10}$ cm$^{-2}$ cumulative fluence.
The chip will be installed in two CMS detectors in the HL-LHC phase, the Barrel Timing Detector (BTL) and the High Granularity Calorimeter (HGCAL).
\end{abstract}


\end{frontmatter}


\section{Introduction}

Starting after Long Shutdown 3, the Large Hadron Collider (LHC) at CERN will enter the so-called high-luminosity phase (HL-LHC).
The maximum instantaneous luminosity will be increased by a factor of 10, hence all the experiments operating at LHC will require a substantial re-design of their detectors to cope with the increased number of collisions, which translates into increased pile-up and radiation levels.

Two new detectors of the CMS experiment will adopt arrays of silicon photomultipliers (SiPMs, also known as multi-pixel photon counters, MPPCs) to detect the light produced in scintillators.
These two detectors are the Barrel Timing Layer (BTL) \cite{CERN-LHCC-2019-003}, a minimum-ionizing particle timing detector in the barrel part of the experiment, and the High Granularity Calorimeter (HGCAL) \cite{CERN-LHCC-2017-023}, in the outer region of the endcap.

The use of SiPMs in highly radioactive environments is particularly challenging since radiation damage induces a sharp increase in the dark count rate (DCR) up to several GHz per device at the detector end-of-life, and a shift in the breakdown voltage that adds to the intrinsic spread between different samples. 
Devices are thus usually cooled at low temperatures to minimize the effect of DCR \cite{CALVI2019243}.
In these difficult working conditions, it is essential for the overall performance of the detector -- especially when aiming at tens of ps of timing resolution like in the BTL -- that the bias voltage is adequately filtered and precisely regulated to the required value throughout the detector lifetime.
The high currents involved when several SiPMs are biased in parallel and the dense packing of these detectors are two other important aspects that encourage the adoption of a point-of-load regulation of the bias supply, close to where the power is dissipated.
Moreover, it is required to perform I-V curves and periodically characterize the SiPM response during the detector lifetime to set each SiPM array's working point as the radiation damage accumulates.

Another need that is common to these detectors is to provide additional low-voltage supplies to the front-end chips with increased filtering, lower noise, and better stability than those provided by DCDC regulators.

The ALDO2 chip was specifically designed to address all these issues for the BTL and HGCAL detectors, and it is the first of this kind in the HEP electronics community.

The device must withstand high radiation levels, especially in BTL, where the radiation will reach a TID of 3.2 Mrad, a 1-MeV-equivalent neutron fluence of $1.9\times10^{14}$ cm$^{-2}$, and a charged hadron fluence of $1.5\times10^{13}$ cm$^{-2}$ at the end of its life.

In the following, we will describe the chip design and its features, together with selected measurements to demonstrate its performance.

\section{Chip description}

The chip is designed in I3T80 0.35~$\mu$m HV CMOS technology from onsemi, which provides both PMOS and NMOS transistors with drain-to-source tolerance of up to 70~V.
This technology was already used in the past by other projects at CERN, and its radiation hardness has been successfully qualified up to the radiation levels required by this application~\cite{FaccioHVradhard}.
A previous version of the chip, ALDO1~\cite{ALDO1}, was designed in a different $0.35\ \mu$m CMOS technology limited to 3.3~V and provided the basis for the design of ALDO2.

\begin{figure}
	\centering
	\includegraphics[width=\linewidth]{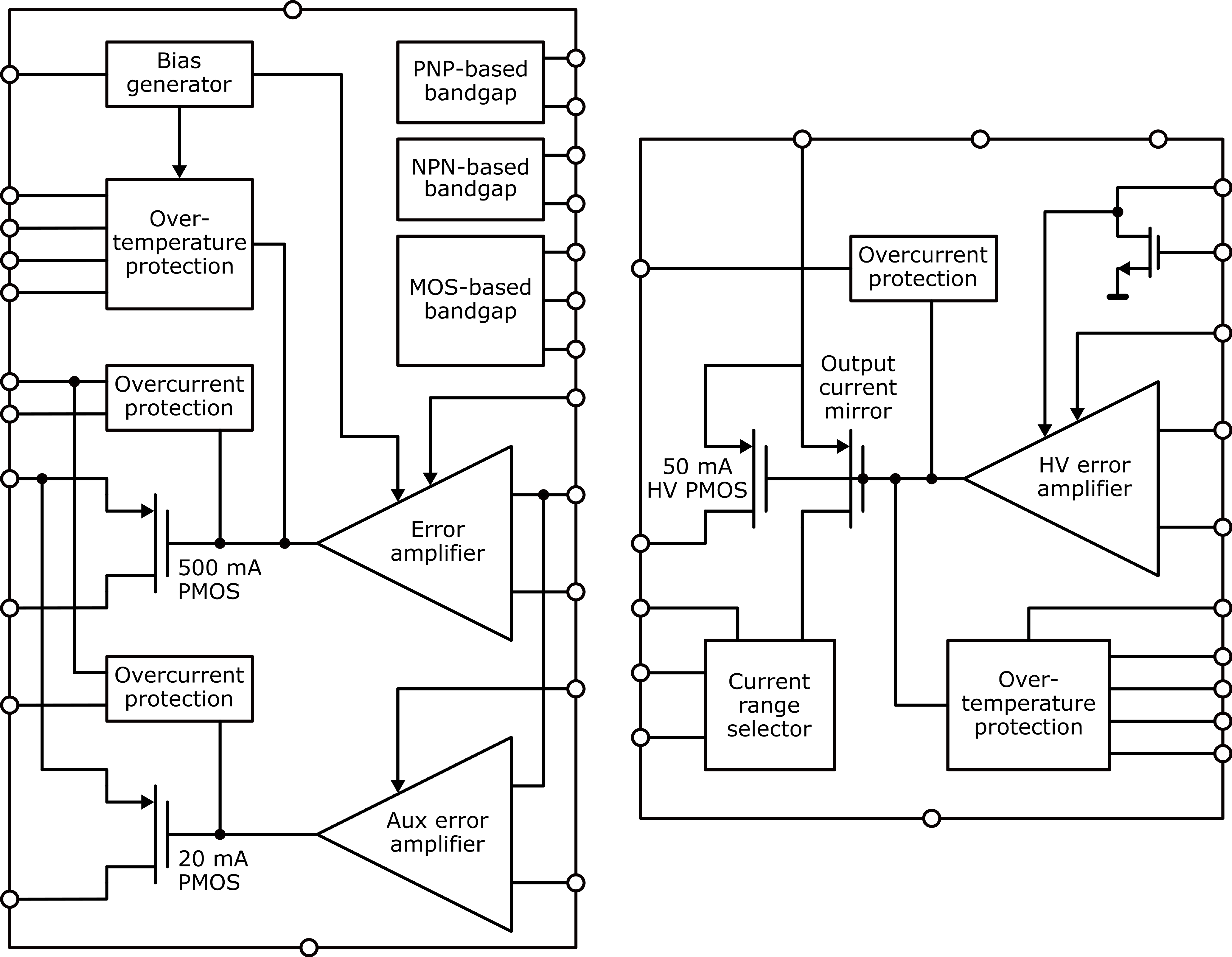}
	\caption{Block diagram of the ALDO2 ASIC. On the \textit{left}: low voltage part of the chip. On the \textit{right}: one of the two (identical) high voltage regulators.}
	\label{fig:blockschematic}
\end{figure}

The ALDO2 chip features four independent and fully adjustable low dropout (LDO) linear regulator channels.
They are all based on the standard LDO topology with an error amplifier driving a PMOS pass transistor that provides the current drive capability to the load.
Figure \ref{fig:blockschematic} shows the ASIC block diagram, on the left the low voltage regulators, on the right the HV ones.

The first regulator provides the main power to front-end chips and thus features low voltage (max 3.3~V input) and high current (max 0.6~A).
The dropout can be as low as 550~mV even at full load and after irradiation up to the levels expected in these detectors.
The second regulator is low voltage and low current (max 20~mA) and can be used to power sensitive circuitry of the front-end chip that requires a separate supply, like internal voltage references.
This regulator is also used internally to generate auxiliary voltage references, like those for the over-temperature protection circuits.

The voltage reference of these two regulators can be selected among three internal bandgap references, one based on PNP bipolar transistors, one on NPN, and one on dynamic-threshold MOSFETs (DTMOS).
In the final application, the chip will use the one that offers the best compromise between radiation hardness and stability.

Both regulators are adjustable, and the user can then set the desired output voltage with the external feedback network from the output to the feedback node of the error amplifier.

The last two regulators, out of the four included in the ALDO2, are those used to generate the bias for SiPM arrays.
They are two identical high voltage regulators with a maximum input voltage of 70~V and a maximum output current of 45~mA.
They allow complete flexibility to the user by tuning both the gain (through the external resistive divider, like for the LV regulators) and the reference voltage (using an external DAC).
In this way, the user would choose a custom, but fixed, output voltage range with the voltage divider and then tune the output voltage during detector operation by programming different values of the reference with the DAC.
In BTL this DAC is implemented in the front-end chip (TOFHIR)~\cite{tofhir}, while in HGCAL the DAC is implemented in the slow control ASIC (GBT-SCA)~\cite{gbtsca}.
The bias voltage regulators also provide a measurement of the output currents, allowing to periodically perform I-V scans and thus determine the breakdown voltage of each array.
This measurement is performed with two ranges ($\times$20 ratio) by mirroring the output current and converting it to a voltage with a precise external resistor.

All the regulators are protected against over-current and over-temperature to avoid self-damage in a faulty condition of the load.
Furthermore, the three main regulators (the LV high-current one and the HV ones) can be individually disabled using external digital signals, allowing power cycling of front-end chips or disabling faulty SiPM arrays.
For BTL, for example, this last feature limits the propagation of the fault of one single SiPM to just one array of 16 SiPMs, rather than to all the 768 SiPMs that share the same bias line.
Unlike readout node trimming solutions, the regulation on the high side of the SIPM bias allows good filtering of noise coming from the supply and does not add parasitics to the input of the readout chip.

The compensation for all the four regulators is external, using low ESR tantalum capacitors.
The BTL detector will adopt low profile (1.8~mm) 50~V ones, given the tight space constraints.

In addition to the choice of onsemi I3T80 technology, radiation hardness is also improved with specific design choices: enclosed layout NMOS transistors (ELTs), ``guarded'' HV NMOS transistors, use of guard-rings and large spacing, abundant substrate contacts, strict anti-latchup rules. 

\begin{figure}
	\centering
	\includegraphics[width=\linewidth]{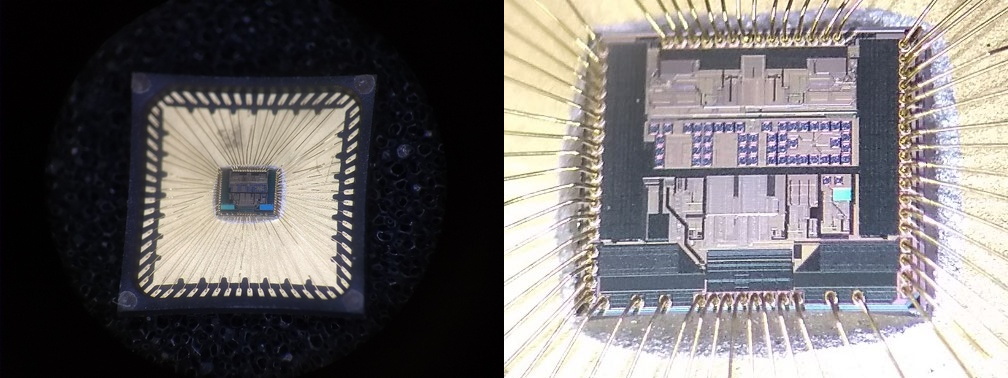}
	\caption{Photos of the ALDO2 die inside an OpenPak QFN64 package.}
	\label{fig:photos}
\end{figure}

The chip die dimension is about $2.5\times2\ \mathrm{mm}^2$ and it is packaged in QFN64 $9\times9\ \mathrm{mm}^2$. Figure \ref{fig:photos} shows two photographs of the ALDO2 die inside an OpenPak QFN.

The chip prototyping phase took three iterations. ALDO2v0, produced in 2019, was the first prototype but already implemented all the chip functionalities.
It was produced in 80 samples and packaged in OpenPak QFN.
The second iteration, ALDO2v1, had a few minor improvements related to over-current limits, bandgap trimming, and over-temperature protection circuit.
About 1000 samples were produced from 8 different wafers, which allowed to test the wafer-by-wafer spread.
ALDO2v1 was packaged in the final molded plastic (punched) QFN64 by ASE.
The last iteration, ALDO2v2, is the one that was sent for production.
The only change was an optional (thereby fail-safe) feature to generate the bias voltages of the HV regulators internally.

\section{Measurements}

The ASIC's performance has been thoroughly characterized on stand-alone test boards and when mounted on the prototype boards designed for BTL and HGCAL detectors that integrate front-end chips and the final SiPM arrays.
These boards were also used successfully during test beam campaigns, proving the ALDO2 functionalities on the field.

In the following, the main results will be presented.

\subsection{Thermal stability and spread}

\begin{figure}
	\centering
	\includegraphics[width=.9\linewidth]{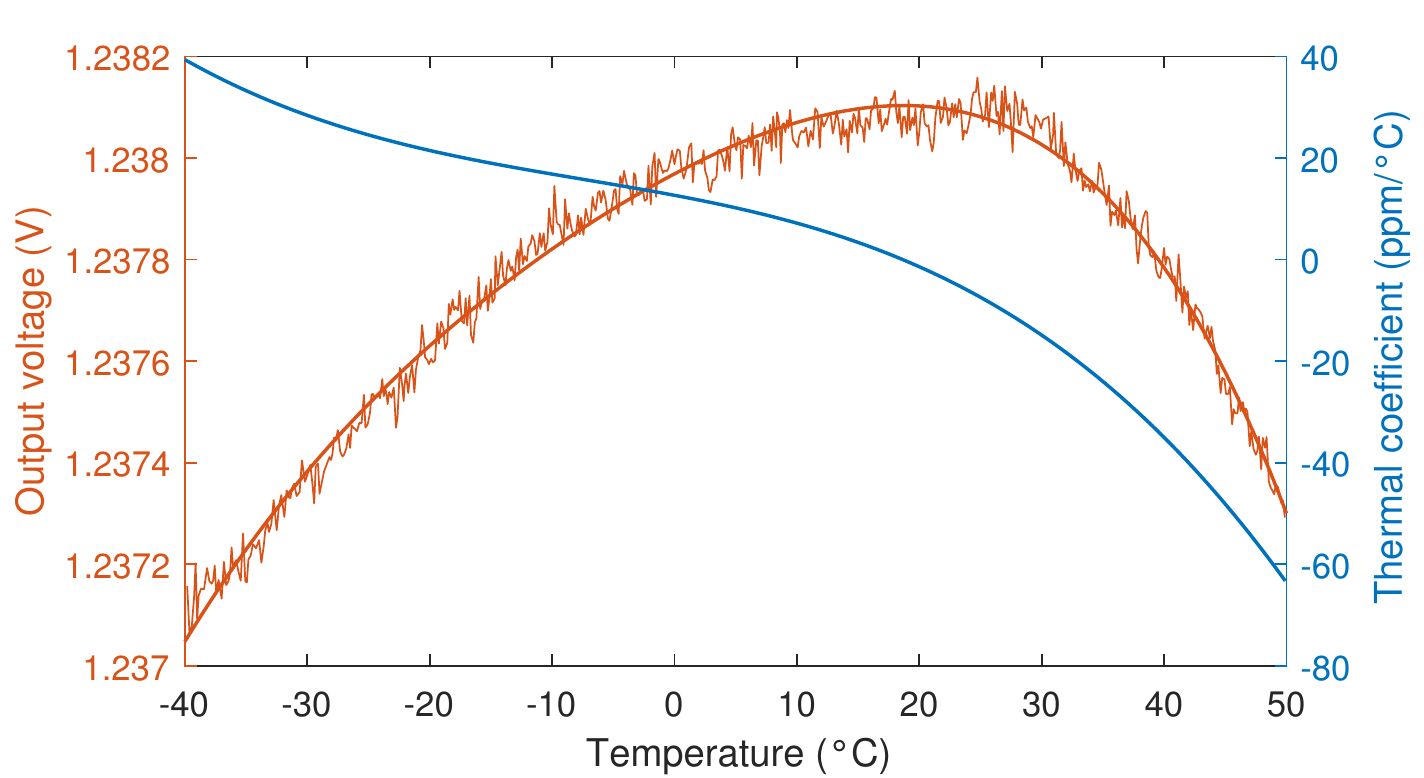}
	\caption{Thermal drift of the main output voltage between $\mathrm{-40\ ^\circ C}$ and $\mathrm{50\ ^\circ C}$.}
	\label{fig:drift}
\end{figure}

The thermal stability of the bandgap voltage references was tested in a climatic chamber.
The PNP-based bandgap showed the best stability, about $\mathrm{20\ ppm/^\circ C}$ from $\mathrm{-40\ ^\circ C}$ to $\mathrm{80\ ^\circ C}$, with flat slope around $\mathrm{0\ ^\circ C}$, as expected.
The DTMOS-based one was the second-best, with about $\mathrm{50\ ppm/^\circ C}$ drift and a higher quadratic slope.
The NPN-based shows a drift of about $\mathrm{120\ ppm/^\circ C}$.
The output voltage shows the same stability of the bandgap used to provide the reference to the error amplifier, confirming that no additional drifts are introduced by the error amplifier or by the pass transistor.
Figure \ref{fig:drift} shows the thermal drift of the output voltage in the range from $\mathrm{-40\ ^\circ C}$ to $\mathrm{50\ ^\circ C}$, using the DTMOS bandgap as reference.
The curve is fitted with a 4th-grade polynomial and then differentiated to get the thermal coefficient as a function of temperature (right axis).

\begin{figure}
	\centering
	\includegraphics[width=.9\linewidth]{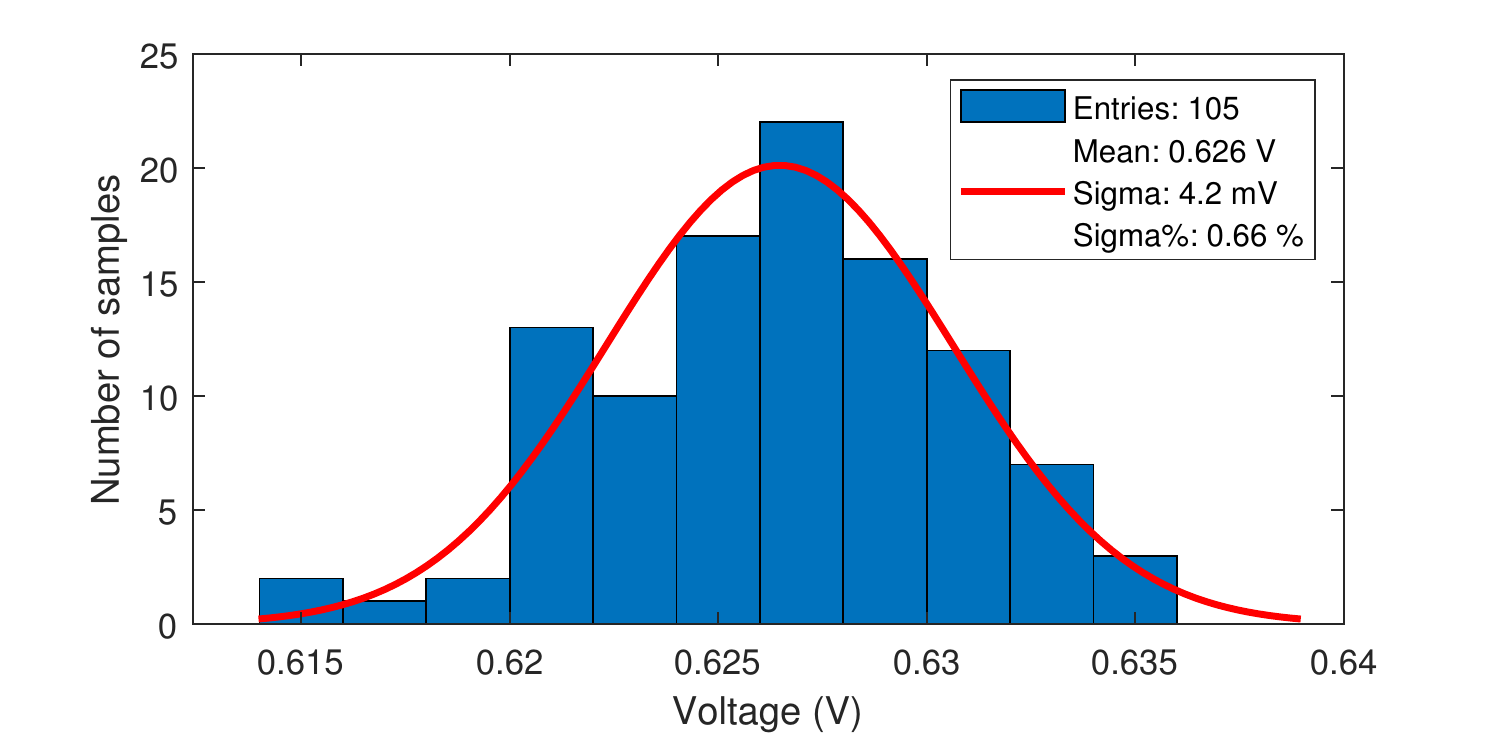}
	\caption{Distribution of the DTMOS-based bandgap reference voltage in the 105 chips tested from the ALDO2v1 prototyping run.}
	\label{fig:spread}
\end{figure}

The spread of the bandgap voltages, and consequently the outputs, is about 0.7\% RMS even with chips coming from different wafers.
Thus there will be no need to trim the value of the bandgap chip by chip.
Figure \ref{fig:spread} shows the spread of the DTMOS-based bandgap voltage in the 105 ALDO2v1 tested.
The spread of the other bandgaps is similar to the one shown.

\subsection{Line regulation, PSRR, and load regulation}

Another important parameter of the ALDO2 is the capability to maintain the output stable, regardless of any input voltage variation, both in DC (line regulation) and over frequency (power supply rejection ratio, PSRR).
These parameters largely depend on the output load and dropout, and are also sensitive to the radiation hardness, as will be shown.

\begin{figure}
	\centering
	\includegraphics[width=.9\linewidth]{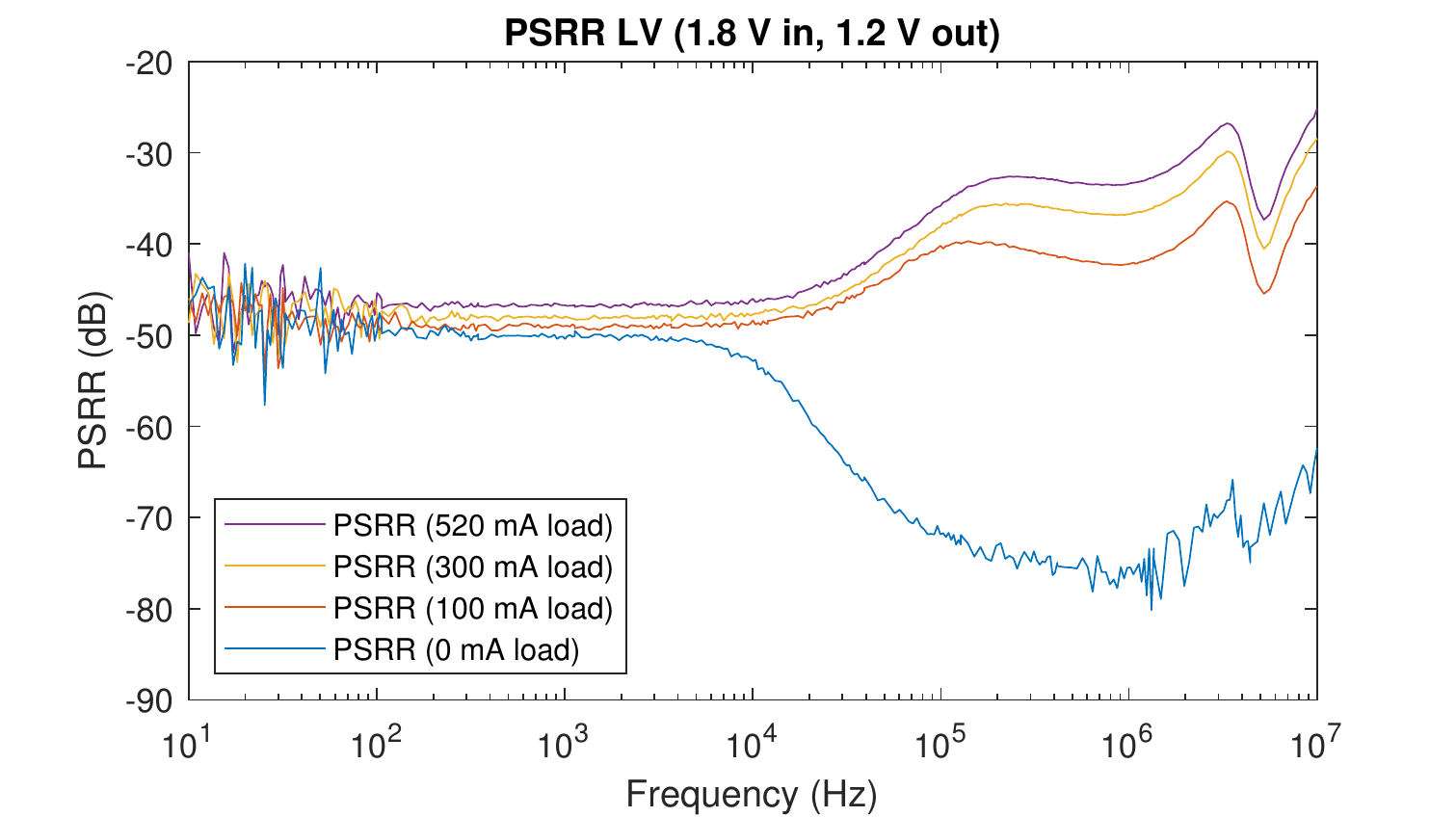}
	\caption{PSRR of the main low voltage regulator at different loads.}
	\label{fig:psrr}
\end{figure}

On the LV regulator, line regulation is about $\mathrm{-46\ dB}$ at 500~mA load, and PSRR is better than $\mathrm{-26\ dB}$ over the entire frequency range up to 10~MHz, as shown in Figure \ref{fig:psrr}, with a dropout of 600~mV.
The HV regulator has a line regulation of $\mathrm{-48\ dB}$ at a load of 43~mA and 2~V dropout.

\begin{figure}
	\centering
	\includegraphics[width=.9\linewidth]{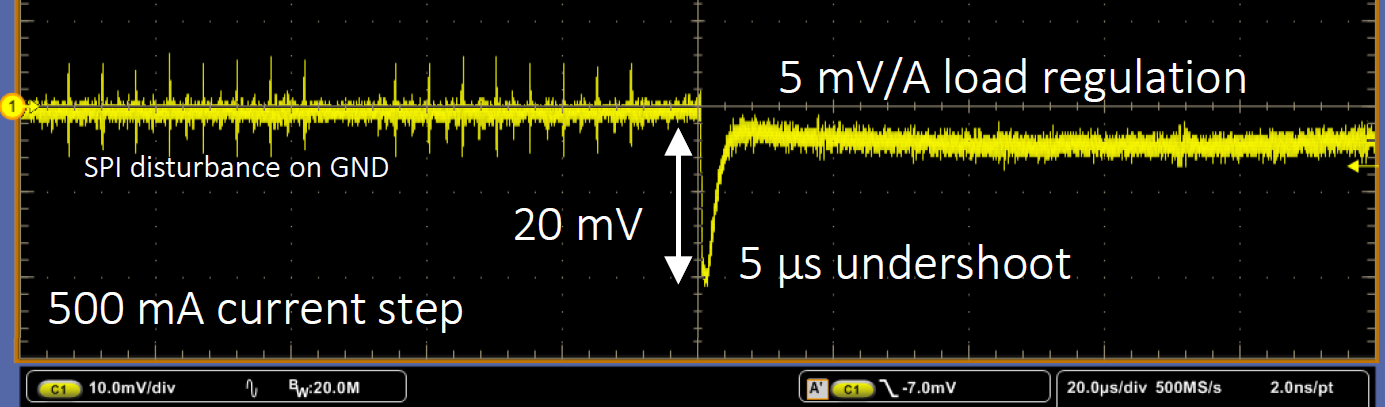}
	
	\vspace*{2mm}
	\includegraphics[width=.9\linewidth]{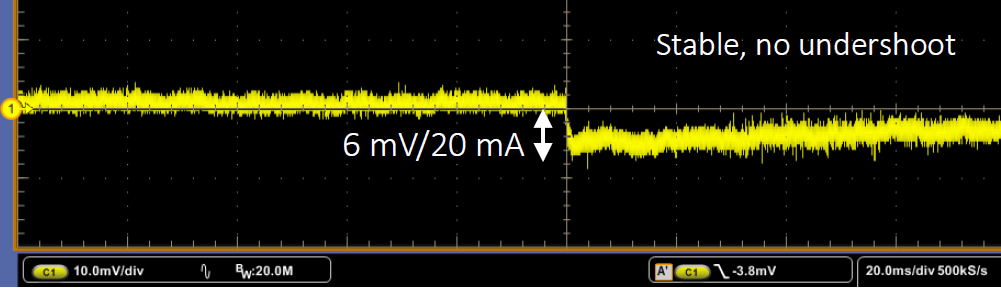}
	\caption{Load current steps on LV (top) and HV (bottom) outputs. The steps are 500~mA on the LV and 20~mA on the HV. Some ground noise on the oscilloscope probe is visible due to the SPI communication to set the output load.}
	\label{fig:step}
\end{figure}

The load regulation (the output voltage variation with different output loads) is 5~mV/A (0.42 \%/A with 1.2~V output) for the LV regulator and 0.3~mV/mA (7.5~ppm/mA with 40~V output) for the HV regulator.
Figure \ref{fig:step} shows two oscilloscope screenshots with load steps on LV (top) and HV (bottom) output, where it is also possible to appreciate the high stability of the regulators since little or no ringing is present.

\subsection{Noise}

ALDO2 is able to effectively filter any input noise thanks to its high PSRR.
Its intrinsic noise at the outputs is $27\ \mu V$ RMS for the LV regulator and $350\ \mu V$ RMS for the HV regulator (at 500~mA and 20~mA load, respectively), in a bandwidth between 10~Hz and 100~MHz.
In the detectors, the main contribution to the HV output voltage noise will come from the DAC reference noise, which cannot be fully filtered due to board space constraints.
The DAC noise is amplified by a factor of about 40 to generate the typical SiPM bias voltage, resulting in an output RMS noise of less than 2~mV RMS.

\section{Radiation hardness tests}

The highest radiation levels that the ALDO2 will have to experience during its operation are those in the BTL detector: a TID of 3.2~Mrad, a 1-MeV-equivalent neutron fluence of $\mathrm{1.9\times10^{14}\ cm^{-2}}$, and a charged hadron fluence of $\mathrm{1.5\times10^{13}\ cm^{-2}}$ after the nominal integrated luminosity of $\mathrm{3000\ fb^{-1}}$.
The HGCAL, although being in the endcap sector, is in fact much further from the interaction point and most of the radiation is already shielded by the inner part of the calorimeter itself.

The radiation hardness qualification was performed in several laboratories.
The TID irradiation with X-rays was done at the Karlsruhe Institute of Technology (Germany) in 2 steps, one at 7~Mrad and one at 20~Mrad, which correspond to a factor 2 and 6 above the expected levels.
The neutron irradiation was performed at LENA nuclear reactor in Pavia (Italy), with 1-MeV-equivalent fluences of $\mathrm{2.5\times10^{14}\ cm^{-2}}$ and $\mathrm{1\times10^{15}\ cm^{-2}}$ ($\times1.3$ and $\times5$).
The heavy-ion irradiation for single-event effects was done at the SIRAD facility in Legnaro (Italy), using several ion types with LETs up to $\mathrm{40\ MeV\ cm^2\ mg^{-1}}$ and a cumulative fluence of $\mathrm{10^{10}\ cm^{-2}}$.

\begin{figure}
	\centering
	\includegraphics[width=.9\linewidth]{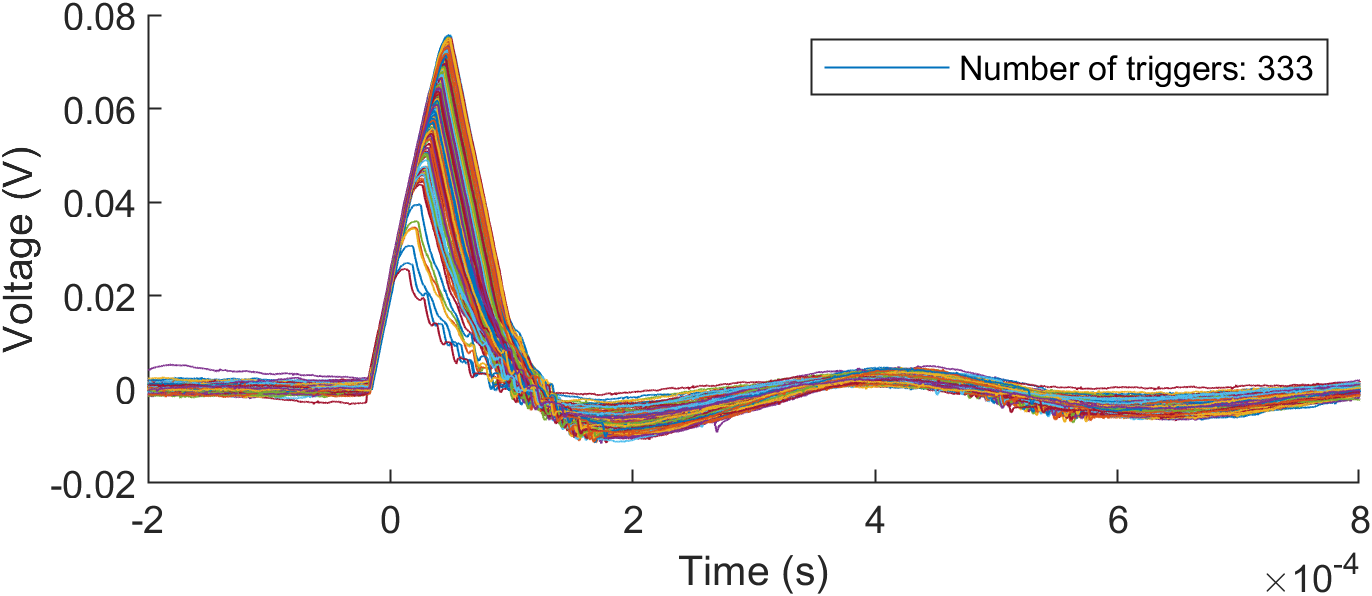}
	\caption{Single-event transients on the output of the HV regulator (during the irradiation with a LET of $\mathrm{28\ MeV\ cm^2\ mg^{-1}}$).}
	\label{fig:set}
\end{figure}

The chips were monitored online during the X-ray and heavy-ion irradiations to check any drift or transient at nominal operating loads.
Single event transients (SETs) were observed at high LET ($\mathrm{28\ MeV\ cm^2\ mg^{-1}}$ and above), initially with an estimated cross-section of $\mathrm{10^{-5}\ cm^{-2}}$.
The cause was then identified, and mitigation was put in place by improving the filtering of the over-temperature protection threshold.
The cross-section of these SETs was thus lowered to $\mathrm{2\times10^{-7}\ cm^{-2}}$, which would correspond to a rate below 1~mHz in the final detector.
It is also worth noting that the amplitude and duration of these transients are very small, 20~mV and 10~$\mu$s for the LV regulator and 80~mV and 100~$\mu$s for the HV regulator, and they are not expected to affect the detector performance.
Figure \ref{fig:set} shows an example of SETs on the HV output voltage.

Measurements on the bandgaps of irradiated devices confirmed that the one that offers the best compromise between radiation hardness and stability is the one based on DTMOS, with drifts below 2\% even after 7~Mrad of TID.
Measurements of the line and load regulation of the irradiated devices allowed to set the minimum dropout to 550~mV for the LV regulator and 3~V for the HV regulator at nominal loads. 

\section{Conclusions and prospects}

The chip development has been completed and the measured performance has fulfilled the requirements of BTL and HGCAL detectors. 
The production of the 19k chips required will be completed by mid-2022.
Due to the selling of the I3T80 onsemi fab in Belgium, where all the prototypes were manufactured, there will not be another possibility to produce more chips in this fab, which demonstrated adequate radiation tolerance.
It was thus decided to manufacture many more chips than needed (45k), just to be prepared in case of unexpectedly low production or packaging yields.
However, a second I3T80 fab is still operating in the USA, although using this one would require to re-do all the radiation hardness qualification, which is not acceptable for the schedule of BTL and HGCAL, but could be pursued if other users would be interested in adopting the ALDO2 ASIC in their detectors.

\section*{Acknowledgments}

The authors would like to thank all the colleagues at KIT (Karlsruhe), LENA (Pavia), and LNL (Legnaro) for their invaluable help and support during the irradiation campaigns.

\bibliography{\jobname}

\end{document}